\documentclass[aps,prl,twocolumn,superscriptaddress,showpacs,preprintnumbers,amsmath,amssymb]{revtex4}
\usepackage{graphicx,color}% Include figure files
\usepackage{dcolumn}% Align table columns on decimal point
\usepackage{bm}% bold math

\begin{document}

%%%%%%%%%%%%%%%%%%%%%%%%
% User definition file
%%%%%%%%%%%%%%%%%%%%%%%%

% defs.tex
%%%%%%%%%%%%%%%%%%%%%%
% New commands
%%%%%%%%%%%%%%%%%%%%%%
%\input{defs-blife}
\def\bz{{B^0}}
\def\bzb{{\overline{B}{}^0}}
\def\bp{{B^+}}
\def\bm{{B^-}}
\def\kl{K_L^0}
\def\dE{{\Delta E}}
\def\mb{{M_{\rm bc}}}
\def\Dt{\Delta t}
\def\Dz{\Delta z}
\def\fol{f_{\rm ol}}
\def\fsig{f_{\rm sig}}
\newcommand{\sinbb}{{\sin2\phi_1}}
 
\newcommand{\ra}{\rightarrow}
\newcommand{\myindent}{\hspace*{2cm}}  % My indent for equation environments
\newcommand{\fCP}{f_{\rm sig}}
\newcommand{\zCP}{z_{\rm sig}}
\newcommand{\tCP}{t_{\rm sig}}
\newcommand{\ftag}{f_{\rm tag}}
\newcommand{\ttag}{t_{\rm tag}}
\newcommand{\ztag}{z_{\rm tag}}
\newcommand{\cala}{{\cal A}}
\newcommand{\calb}{{\cal B}}
\newcommand{\cals}{{\cal S}}
\newcommand{\dm}{\Delta m_d}
\newcommand{\dmd}{\dm}
\def\taubz{{\tau_\bz}}
\def\taubp{{\tau_\bp}}
\def\ks{{K_S^0}}
\newcommand{\btosqq}{b \to s\overline{q}q}
\newcommand{\btosss}{b \to s\overline{s}s}
\newcommand*{\dwl}{\ensuremath{{\Delta w_l}}}
\newcommand*{\fq}{\ensuremath{q}}
\def\kz{{K^0}}
\def\kp{{K^+}}
\def\km{{K^-}}
\def\fzero{{f_0(980)}}
\def\pip{{\pi^+}}
\def\pim{{\pi^-}}
\def\piz{{\pi^0}}
\def\kstar{{K^\ast}}
\def\kstarz{{K^{\ast 0}}}
\def\kstarp{{K^{\ast +}}}
\def\kstarm{{K^{\ast -}}}
\def\kstarpm{{K^{\ast\pm}}}
\def\kl{{K_L^0}}
\def\bbar{{\overline{B}}}
\def\ufs{{\Upsilon(4S)}}
\def\nev{{N_{\rm ev}}}
\def\nsig{{N_{\rm sig}}}
\def\Nev{\nev}
\def\nsigmc{{N_{\rm sig}^{\rm MC}}}
\def\nbkg{{N_{\rm bkg}}}

\def\jpsi{{J/\psi}}
\def\dminus{{D^-}}
\def\dplus{{D^+}}
\def\dsm{{D^{*-}}}
\def\dpm{{D^{*+}}}
\def\rhop{{\rho^+}}
\def\rhom{{\rho^-}}
\def\rhoz{{\rho^0}}
\def\dzero{{D^0}}
\def\dzerob{{\overline{D}{}^0}}

\def\dzb{{\overline{D}{}^0}}
\newcommand{\dslnu}{D^{*-}\ell^+\nu}
\newcommand{\bzdslnu}{\bz \to \dslnu}
\newcommand*{\fflv}{\ensuremath{{f_\textrm{flv}}}}
\newcommand{\thetabdl}{\theta_{B,D^*\ell}}
\newcommand{\cosbdl}{\cos\thetabdl}

\def\sperp{{S_{\perp}}}
\def\lsig{{\cal L}_{\rm sig}}
\def\lbkg{{\cal L}_{\rm bkg}}
\def\rsigbkg{{\cal R}_{\rm s/b}}
\def\calf{{\cal F}}
\def\rkpi{{\cal R}_{K/\pi}}

\def\mgg{M_{\gamma\gamma}}
\def\ppizcms{p_\piz^{\rm cms}}
\def\ppizlab{p_\piz^{\rm lab}}
\def\ppipcms{p_\pip^{\rm cms}}

\def\pbstar{p_B^{\rm cms}}

\newcommand*{\eeff}{\ensuremath{\epsilon_\textrm{eff}}}
\def\egcms{{E_\gamma^{\rm cms}}}

\def\lnsig{{\cal L}_{N_{\rm sig}}}
\def\lzero{{\cal L}_0}

\def\acpraw{{A_{CP}^{\rm raw}}}

\def\qq{q\bar{q}}

\def\GeV{\,{\rm GeV}}
\def\GeVc{\,{\rm GeV}/c}
\def\GeVcc{\,{\rm GeV}/c^2}
\def\MeVcc{\,{\rm MeV}/c^2}

\newcommand{\BaBar}{{\sl BABAR}}
\def\Rsig{R_{\rm sig}}
\def\Rbkg{R_{\rm bkg}}
%\input{defs.tex}

% sss04pqx_defs.tex
%%%%%%%%%%%%%%%%%%
% for 3Ks states
%%%%%%%%%%%%%%%%%%
\def\kspm{{K_S^{+-}}}
\def\kszz{{K_S^{00}}}

%%%%%%%%%%%%%%%%%%%%%%%%
% flavor tagging for SVD2
%%%%%%%%%%%%%%%%%%%%%%%%
\def\efftot{{0.30\pm 0.01}}

%%%%%%%%%%%%%%%%%%%%%%%%
% sin2phi1 world average
%%%%%%%%%%%%%%%%%%%%%%%%
%\def\sinbbWA{+0.736}
%\def\sinbbERR{0.049}
\def\sinbbWA{+0.726}
\def\sinbbERR{0.037}
\def\sinbbWAResult{\sinbbWA\pm\sinbbERR}

%%%%%%%%%%%%%%%%%%%%%%%%
% Previous results
%%%%%%%%%%%%%%%%%%%%%%%%
\def\SphiksResPrv{-0.96\pm0.50^{+0.09}_{-0.11}}
\def\AphiksResPrv{-0.15\pm0.29\pm0.07}
\def\SetapksResPrv{+0.43\pm0.27\pm0.05}
\def\AetapksResPrv{-0.01\pm0.16\pm0.04}
\def\SkpkmksResPrv{-0.51\pm0.26\pm0.05}
\def\AkpkmksResPrv{-0.17\pm0.16\pm0.04}

%%%%%%%%%%%%%%%%%%%%%%%%%%%%
% New results in this paper
%%%%%%%%%%%%%%%%%%%%%%%%%%%%

\def\NBevksksks{167} 
\def\PBksksks{0.53} \def\NBsigksksks{88\pm 13}

\def\NBevkspmkspmkspm{128} 
\def\PBkspmkspmkspm{0.56} \def\NBsigkspmkspmkspm{72\pm 10}

\def\NBevkspmkspmkszz{39} 
\def\PBkspmkspmkszz{0.40} \def\NBsigkspmkspmkszz{16\pm 8}

\def\NAevksksks{117} 
\def\PAksksks{0.54} \def\NAsigksksks{63\pm 9}

\def\NAevkspmkspmkspm{96} 
\def\PAkspmkspmkspm{0.56} \def\NAsigkspmkspmkspm{54\pm 8}

\def\NAevkspmkspmkszz{21} 
\def\PAkspmkspmkszz{0.40} \def\NAsigkspmkspmkszz{8\pm 4}

\def\Nevkspizgm{227} 
\def\Pkspizgm{0.50} \def\Nsigkspizgm{105 \pm 14}

\def\SksksksVal{+1.26} \def\SksksksStat{0.68} \def\SksksksSyst{0.18}
\def\AksksksVal{+0.54} \def\AksksksStat{0.34} \def\AksksksSyst{0.08}

\def\SkspizgmVal{-0.58} \def\SkspizgmStat{^{+0.46}_{-0.38}} \def\SkspizgmSyst{0.11}
\def\AkspizgmVal{+0.03} \def\AkspizgmStat{0.34} \def\AkspizgmSyst{0.11}

\def\SksksksResult{\SksksksVal\pm\SksksksStat\pm\SksksksSyst}
\def\SksksksResultSS
  {\SksksksVal\pm\SksksksStat\mbox{(stat)}\pm\SksksksSyst\mbox{(syst)}}

\def\AksksksResult{\AksksksVal\pm\AksksksStat\pm\AksksksSyst}
\def\AksksksResultSS
  {\AksksksVal\pm\AksksksStat\mbox{(stat)}\pm\AksksksSyst\mbox{(syst)}}

\def\SkspizgmResult{\SkspizgmVal\SkspizgmStat\pm\SkspizgmSyst}
\def\SkspizgmResultSS
  {\SkspizgmVal\SkspizgmStat\mbox{(stat)}\pm\SkspizgmSyst\mbox{(syst)}}

\def\AkspizgmResult{\AkspizgmVal\pm\AkspizgmStat\pm\AkspizgmSyst}
\def\AkspizgmResultSS
  {\AkspizgmVal\pm\AkspizgmStat\mbox{(stat)}\pm\AkspizgmSyst\mbox{(syst)}}

\def\SbsqqNewVal{+0.39} \def\SbsqqNewErr{0.11}
\def\SbsqqNewResult{\SbsqqNewVal\pm\SbsqqNewErr}

%%%%%%%%%%%%%%%%%%%%%%%%
% ICHEP04 Results
%%%%%%%%%%%%%%%%%%%%%%%%
\def\Nevjpsikspm{xxxx} 
\def\Pjpsikspm{0.xx} \def\Nsigjpsikspm{xxxx\pm xx}
 \def\Nevjpsikszz{xxx} 
\def\Pjpsikszz{0.xx} \def\Nsigjpsikszz{xxx\pm xx} 
  \def\Nevjpsikl{xxxx}     
  \def\Pjpsikl{0.xx}    \def\Nsigjpsikl{xxxx\pm xx}
    \def\Nevphiks{221}       
   \def\Pphiks{0.63}     \def\Nsigphiks{139 \pm 14}
    \def\Nevphikl{207}       
   \def\Pphikl{0.17}     \def\Nsigphikl{36 \pm 15}
\def\Nevkpkmks{718}     
  \def\Pkpkmks{0.56}    \def\Nsigkpkmks{399\pm 28}
\def\NevkspizH{298}      
   \def\PkspizH{0.55}     \def\NsigkspizH{168\pm 16}
\def\NevkspizL{499}      
   \def\PkspizL{0.17}     \def\NsigkspizL{83\pm 18}
\def\Nevkstarzgm{92} 
\def\Pkstarzgm{0.65}  \def\Nsigkstarzgm{57 \pm 9}
\def\Nevetapks{842}   
  \def\Petapks{0.61}    \def\Nsigetapks{512 \pm 27}
\def\Nevomegaks{56}  
 \def\Pomegaks{0.56}    \def\Nsigomegaks{31 \pm 7}
\def\Nevfzeroetcks{178}  
 \def\Pfzeroetcks{0.58}    \def\Nsigfzeroetcks{102 \pm 12}
 \def\Pfzeroks{0.53}    \def\Nsigfzeroks{94 \pm 14}
%%   Cf. 0.53 = 0.58 x 0.912, where 0.912 is the
%%       f0 Ks fraction obtained by a fit to the pi+pi- 
%%       invariant distribution.
%
%%%%%%%%%%%%%%%%%%%%%%%%%%%%%%%%%%%%%%%%%%%%%%%%%%%%%%%%%%%%%%%%%%%%%%%%%%%
\def\SjpsikzVal{+0.666}   \def\SjpsikzStat{0.046}   \def\SjpsikzSyst{x.xx}
\def\AjpsikzVal{+0.023}   \def\AjpsikzStat{0.031}   \def\AjpsikzSyst{x.xx}
\def\SphikzVal{+0.06}    \def\SphikzStat{0.33}    \def\SphikzSyst{0.09}
\def\AphikzVal{+0.08}    \def\AphikzStat{0.22}    \def\AphikzSyst{0.09}
\def\SphiksVal{-x.xx}    \def\SphiksStat{x.xx}    \def\SphiksSyst{x.xx}
\def\AphiksVal{-x.xx}    \def\AphiksStat{x.xx}    \def\AphiksSyst{x.xx}
\def\SphiklVal{-x.xx}    \def\SphiklStat{x.xx}    \def\SphiklSyst{x.xx}
\def\AphiklVal{-x.xx}    \def\AphiklStat{x.xx}    \def\AphiklSyst{x.xx}
\def\SkpkmksVal{-0.49}   \def\SkpkmksStat{0.18}   \def\SkpkmksSyst{0.04}
\def\AkpkmksVal{-0.08}   \def\AkpkmksStat{0.12}   \def\AkpkmksSyst{0.07}
\def\SkpkmksFCP{^{+0.18}_{-0.00}}
\def\SkspizVal{+0.30}    \def\SkspizStat{0.59}    \def\SkspizSyst{0.11}
\def\AkspizVal{-0.12}    \def\AkspizStat{0.20}    \def\AkspizSyst{0.07}
\def\SkstarzgmVal{-0.79} \def\SkstarzgmStat{^{+0.63}_{-0.50}} \def\SkstarzgmSyst{0.10}
\def\AkstarzgmVal{-0.00} \def\AkstarzgmStat{0.38} \def\AkstarzgmSyst{x.xx}
\def\AkstargmResultNakaoVal{0.015} \def\AkstargmResultNakaoStat{0.044} \def\AkstargmResultNakaoSyst{0.026}
\def\SetapksVal{+0.65}   \def\SetapksStat{0.18}   \def\SetapksSyst{0.04}
\def\AetapksVal{-0.19}   \def\AetapksStat{0.11}   \def\AetapksSyst{0.05}
\def\SomegaksVal{+0.75}  \def\SomegaksStat{0.64}  \def\SomegaksSyst{^{+0.13}_{-0.16}}
\def\AomegaksVal{+0.26}  \def\AomegaksStat{0.48}  \def\AomegaksSyst{0.15}
\def\SfzeroksVal{+0.47}  \def\SfzeroksStat{0.41}  \def\SfzeroksSyst{0.08}
\def\AfzeroksVal{-0.39}  \def\AfzeroksStat{0.27}  \def\AfzeroksSyst{0.08}
\def\SbsqqVal{+0.43} \def\SbsqqErr{^{+0.12}_{-0.11}}

%%%%%%%%%%%%%%%%%%%%%%%%%%%%%%%%%%%%%%%%%%%%%%%%%%%%%%%%%%%%%%%%%%%%%%%%%%
\def\SjpsikzResult{\SjpsikzVal\pm\SjpsikzStat\pm\SjpsikzSyst}
\def\SjpsikzResultSS
  {\SjpsikzVal\pm\SjpsikzStat\mbox{(stat)}\pm\SjpsikzSyst\mbox{(syst)}}
\def\AjpsikzResult{\AjpsikzVal\pm\AjpsikzStat\pm\AjpsikzSyst}
\def\AjpsikzResultSS
  {\AjpsikzVal\pm\AjpsikzStat\mbox{(stat)}\pm\AjpsikzSyst\mbox{(syst)}}
\def\SphikzResult{\SphikzVal\pm\SphikzStat\pm\SphikzSyst}
\def\SphikzResultSS
  {\SphikzVal\pm\SphikzStat\mbox{(stat)}\pm\SphikzSyst\mbox{(syst)}}
\def\AphikzResult{\AphikzVal\pm\AphikzStat\pm\AphikzSyst}
\def\AphikzResultSS
  {\AphikzVal\pm\AphikzStat\mbox{(stat)}\pm\AphikzSyst\mbox{(syst)}}
\def\SphiksResult{\SphiksVal\pm\SphiksStat\pm\SphiksSyst}
\def\SphiksResultSS
  {\SphiksVal\pm\SphiksStat\mbox{(stat)}\pm\SphiksSyst\mbox{(syst)}}
\def\AphiksResult{\AphiksVal\pm\AphiksStat\pm\AphiksSyst}
\def\AphiksResultSS
  {\AphiksVal\pm\AphiksStat\mbox{(stat)}\pm\AphiksSyst\mbox{(syst)}}
\def\SphiklResult{\SphiklVal\pm\SphiklStat\pm\SphiklSyst}
\def\SphiklResultSS
  {\SphiklVal\pm\SphiklStat\mbox{(stat)}\pm\SphiklSyst\mbox{(syst)}}
\def\AphiklResult{\AphiklVal\pm\AphiklStat\pm\AphiklSyst}
\def\AphiklResultSS
  {\AphiklVal\pm\AphiklStat\mbox{(stat)}\pm\AphiklSyst\mbox{(syst)}}
\def\SkpkmksResult{\SkpkmksVal\pm\SkpkmksStat\pm\SkpkmksSyst}
\def\SkpkmksResultSS
  {\SkpkmksVal\pm\SkpkmksStat\mbox{(stat)}\pm\SkpkmksSyst\mbox{(syst)}}
\def\AkpkmksResult{\AkpkmksVal\pm\AkpkmksStat\pm\AkpkmksSyst}
\def\AkpkmksResultSS
  {\AkpkmksVal\pm\AkpkmksStat\mbox{(stat)}\pm\AkpkmksSyst\mbox{(syst)}}
\def\SkspizResult{\SkspizVal\pm\SkspizStat\pm\SkspizSyst}
\def\SkspizResultSS
  {\SkspizVal\pm\SkspizStat\mbox{(stat)}\pm\SkspizSyst\mbox{(syst)}}
\def\AkspizResult{\AkspizVal\pm\AkspizStat\pm\AkspizSyst}
\def\AkspizResultSS
  {\AkspizVal\pm\AkspizStat\mbox{(stat)}\pm\AkspizSyst\mbox{(syst)}}
\def\SkstarzgmResult{\SkstarzgmVal\SkstarzgmStat\pm\SkstarzgmSyst}
\def\SkstarzgmResultSS
  {\SkstarzgmVal\SkstarzgmStat\mbox{(stat)}\pm\SkstarzgmSyst\mbox{(syst)}}
\def\AkstarzgmResult{\AkstarzgmVal\pm\AkstarzgmStat\pm\AkstarzgmSyst}
\def\AkstarzgmResultSS
  {\AkstarzgmVal\pm\AkstarzgmStat\mbox{(stat)}\pm\AkstarzgmSyst\mbox{(syst)}}
\def\AkstargmResultNakao{\AkstargmResultNakaoVal\AkstargmResultNakaoStat\pm\AkstargmResultNakaoSyst}
\def\SetapksResult{\SetapksVal\pm\SetapksStat\pm\SetapksSyst}
\def\SetapksResultSS
  {\SetapksVal\pm\SetapksStat\mbox{(stat)}\pm\SetapksSyst\mbox{(syst)}}
\def\AetapksResult{\AetapksVal\pm\AetapksStat\pm\AetapksSyst}
\def\AetapksResultSS
  {\AetapksVal\pm\AetapksStat\mbox{(stat)}\pm\AetapksSyst\mbox{(syst)}}
\def\SomegaksResult{\SomegaksVal\pm\SomegaksStat\SomegaksSyst}
\def\SomegaksResultSS
  {\SomegaksVal\pm\SomegaksStat\mbox{(stat)}\SomegaksSyst\mbox{(syst)}}
\def\AomegaksResult{\AomegaksVal\pm\AomegaksStat\pm\AomegaksSyst}
\def\AomegaksResultSS
  {\AomegaksVal\pm\AomegaksStat\mbox{(stat)}\pm\AomegaksSyst\mbox{(syst)}}
\def\SfzeroksResult{\SfzeroksVal\pm\SfzeroksStat\pm\SfzeroksSyst}
\def\SfzeroksResultSS
  {\SfzeroksVal\pm\SfzeroksStat\mbox{(stat)}\pm\SfzeroksSyst\mbox{(syst)}}
\def\AfzeroksResult{\AfzeroksVal\pm\AfzeroksStat\pm\AfzeroksSyst}
\def\AfzeroksResultSS
  {\AfzeroksVal\pm\AfzeroksStat\mbox{(stat)}\pm\AfzeroksSyst\mbox{(syst)}}
\def\SbsqqResult{\SbsqqVal\SbsqqErr}

\title{\quad\\[0.5cm] \boldmath Measurement of Time-Dependent 
{\boldmath $CP$}-Violating\\ Asymmetry
in $\bz\to\ks\piz\gamma$ Decay}

\date{\today}% It is always \today, today,

%auth.tex
%%% Paper:    B0 -> KS pi0 gamma TCPV
%%% Journal:  Physical Review Letters
%%% Contacts: Y. Ushiroda (ushiroda@bmail.kek.jp)
%%%           K. Sumisawa (sumisawa@bmail.kek.jp)
%%% Non-responding authors or those who said NO are commented out.
%%% ====================================================================
%%% Click the RELOAD button on your web browser to see the updated file.
%%% ====================================================================
%%% Use \input{author} to insert this material into your latex file.
%%%%% Force institutions to appear in alphabetical order when typeset.

\affiliation{Budker Institute of Nuclear Physics, Novosibirsk}
\affiliation{Chiba University, Chiba}
\affiliation{Chonnam National University, Kwangju}

\affiliation{University of Cincinnati, Cincinnati, Ohio 45221}

\affiliation{Gyeongsang National University, Chinju}
\affiliation{University of Hawaii, Honolulu, Hawaii 96822}
\affiliation{High Energy Accelerator Research Organization (KEK), Tsukuba}

\affiliation{Institute of High Energy Physics, Chinese Academy of Sciences, Beijing}
\affiliation{Institute of High Energy Physics, Vienna}
\affiliation{Institute for Theoretical and Experimental Physics, Moscow}
\affiliation{J. Stefan Institute, Ljubljana}
\affiliation{Kanagawa University, Yokohama}
\affiliation{Korea University, Seoul}

\affiliation{Kyungpook National University, Taegu}
\affiliation{Swiss Federal Institute of Technology of Lausanne, EPFL, Lausanne}
\affiliation{University of Ljubljana, Ljubljana}
\affiliation{University of Maribor, Maribor}
\affiliation{University of Melbourne, Victoria}
\affiliation{Nagoya University, Nagoya}
\affiliation{Nara Women's University, Nara}
\affiliation{National Central University, Chung-li}

\affiliation{National United University, Miao Li}
\affiliation{Department of Physics, National Taiwan University, Taipei}
\affiliation{H. Niewodniczanski Institute of Nuclear Physics, Krakow}
\affiliation{Nihon Dental College, Niigata}
\affiliation{Niigata University, Niigata}
\affiliation{Osaka City University, Osaka}
\affiliation{Osaka University, Osaka}
\affiliation{Panjab University, Chandigarh}
\affiliation{Peking University, Beijing}
\affiliation{Princeton University, Princeton, New Jersey 08545}

\affiliation{Saga University, Saga}
\affiliation{University of Science and Technology of China, Hefei}
\affiliation{Seoul National University, Seoul}
\affiliation{Sungkyunkwan University, Suwon}
\affiliation{University of Sydney, Sydney NSW}
\affiliation{Tata Institute of Fundamental Research, Bombay}
\affiliation{Toho University, Funabashi}
\affiliation{Tohoku Gakuin University, Tagajo}
\affiliation{Tohoku University, Sendai}
\affiliation{Department of Physics, University of Tokyo, Tokyo}
\affiliation{Tokyo Institute of Technology, Tokyo}
\affiliation{Tokyo Metropolitan University, Tokyo}
\affiliation{Tokyo University of Agriculture and Technology, Tokyo}

\affiliation{University of Tsukuba, Tsukuba}

\affiliation{Virginia Polytechnic Institute and State University, Blacksburg, Virginia 24061}
\affiliation{Yonsei University, Seoul}
   \author{Y.~Ushiroda}\affiliation{High Energy Accelerator Research Organization (KEK), Tsukuba} % KEK
   \author{K.~Sumisawa}\affiliation{Osaka University, Osaka}\affiliation{High Energy Accelerator Research Organization (KEK), Tsukuba} % Osaka
   \author{M.~Hazumi}\affiliation{High Energy Accelerator Research Organization (KEK), Tsukuba} % KEK
   \author{K.~Abe}\affiliation{High Energy Accelerator Research Organization (KEK), Tsukuba} % KEK
   \author{K.~Abe}\affiliation{Tohoku Gakuin University, Tagajo} % TohokuGakuin

   \author{I.~Adachi}\affiliation{High Energy Accelerator Research Organization (KEK), Tsukuba} % KEK
   \author{H.~Aihara}\affiliation{Department of Physics, University of Tokyo, Tokyo} % Tokyo

   \author{Y.~Asano}\affiliation{University of Tsukuba, Tsukuba} % Tsukuba

   \author{V.~Aulchenko}\affiliation{Budker Institute of Nuclear Physics, Novosibirsk} % BINP
    \author{T.~Aushev}\affiliation{Institute for Theoretical and Experimental Physics, Moscow} % ITEP

% ...
% Ljubljana

\collaboration{The Belle Collaboration}

\begin{abstract}
 We present a new measurement of $CP$-violation parameters
 in $\bz\to\ks\piz\gamma$ decay
 based on a sample of $275\times 10^6$ $B\bbar$ pairs
 collected at the $\ufs$ resonance with
 the Belle detector at the KEKB energy-asymmetric $e^+e^-$ collider.
 One of the $B$ mesons 
 is fully reconstructed in
 the $\bz\to\ks\piz\gamma$ decay.
 The flavor of the accompanying $B$ meson is identified from
 its decay products.
 $CP$-violation parameters
 are obtained from the asymmetry in the distribution of
 the proper-time intervals between the two $B$ decays.
 We obtain $\cals_{\ks\piz\gamma}=\SkspizgmResultSS$ and
 $\cala_{\ks\piz\gamma}=\AkspizgmResultSS$,
 for the $\ks\piz$ invariant mass covering the full range up to
 $1.8\GeVcc$. 
  We also measure the $CP$-violation parameters for the case
 $\bz\to\kstarz(\to\ks\piz)\gamma$ and obtain 
 $\cals_{\kstarz\gamma}=\SkstarzgmResultSS$ for $\cala_{\kstarz\gamma}$
 fixed at 0.
\end{abstract}


\maketitle

%%%%%%%%%%%%%%%%%%%%%%%%%%%
% Main Part
%%%%%%%%%%%%%%%%%%%%%%%%%%%
%%%%%%%%%%%%%%%%%%%%%%%
%\section{Introduction}
%\label{sec:introduction}
%%%%%%%%%%%%%%%%%%%%%%%%

In the Standard Model (SM), $CP$ violation arises from an irreducible
phase, the Kobayashi-Maskawa (KM) phase~\cite{Kobayashi:1973fv}, in the
weak-interaction quark-mixing matrix.
% In particular, the SM predicts $CP$ asymmetries in the time-dependent
%rates for $\bz$ and $\bzb$ decays to a common $CP$ eigenstate
%$\fCP$~\cite{bib:sanda}.
The phenomena of time-dependent $CP$ violation in decays through
radiative penguin processes such as $b\to s\gamma$ are sensitive to
physics beyond the SM.
%~\cite{Atwood:1997zr,Atwood:2004jj}.
Within the SM, the photon emitted from a $\bz$ ($\bzb$)
meson is dominantly right-handed (left-handed).
Therefore the polarization of the photon carries
information on the original $b$-flavor and the decay
is, thus, almost flavor-specific.
As a result, the SM predicts 
a small asymmetry~\cite{Atwood:1997zr,Grinstein:2004uu} and
any significant deviation from this expectation
would be a manifestation of new physics.
Recently it was pointed out that 
in decays of the type $\bz\to P^0Q^0\gamma$, where $P^0$ and $Q^0$
represent any $CP$ eigenstate spin-0 neutral particles
(e.g. $P^0 = \ks$ and $Q^0 = \piz$),
many new physics effects on the mixing-induced 
$CP$ violation do not depend on the resonant structure
of the $P^0Q^0$ system~\cite{Atwood:2004jj}.
In this Letter, we describe two measurements of $CP$
asymmetries:
one with a $\ks\piz$ mass range restricted around the
$\kstarz$~\cite{bib:Kstar} mass,
and the other with an extended $\ks\piz$ mass range~\cite{note:Grinstein}.
% based on the new proposal.
Whenever necessary, we distinguish these two by
referring to them as $\kstarz\gamma$ and $\ks\piz\gamma$, respectively;
otherwise the analysis procedure is common to
both measurements since
it was first optimized for
the $\kstarz\gamma$ and was extended for the $\ks\piz\gamma$ later.

At the KEKB energy-asymmetric $e^+e^-$ (3.5 on 8.0$\GeV$)
collider~\cite{bib:KEKB}, the $\Upsilon(4S)$ is produced with a Lorentz
boost of $\beta\gamma=0.425$ along the $z$ axis defined as antiparallel
to the $e^+$ beam direction.
In the decay chain $\Upsilon(4S)\to \bz\bzb \to \fCP \ftag$, where one
of the $B$ mesons decays at time $\tCP$ to a final state $\fCP$, which
is our signal mode, and the other decays at time $\ttag$ to a final
state $\ftag$ that distinguishes between $B^0$ and $\bzb$, the decay
rate has a time dependence given by
\begin{eqnarray}
\label{eq:psig}
{\cal P}(\Dt) = 
\frac{e^{-|\Dt|/{\taubz}}}{4{\taubz}}
\biggl\{1 &+& \fq
\Bigl[ \cals\sin(\dmd\Dt) \nonumber \\
   &+& \cala\cos(\dmd\Dt)
\Bigr]
\biggr\}.
\end{eqnarray}
Here $\cals$ and $\cala$ are $CP$-violation parameters, $\taubz$ is the
$B^0$ lifetime, $\dmd$ is the mass difference between the two $B^0$ mass
eigenstates, $\Dt$ is the time difference $\tCP - \ttag$, and the
$b$-flavor charge $\fq$ = +1 ($-1$) when the tagging $B$ meson is a
$B^0$ ($\bzb$).
Since the $B^0$ and $\bzb$ mesons are approximately at 
rest in the $\Upsilon(4S)$ center-of-mass system (c.m.s.),
$\Dt$ can be determined from the displacement in $z$ 
between the $\fCP$ and $f_{\rm tag}$ decay vertices:
$\Delta t \simeq (\zCP - \ztag)/(\beta\gamma c)
 \equiv \Delta z/(\beta\gamma c)$.

%The Belle detector~\cite{Belle} is a large-solid-angle magnetic
%spectrometer
%that consists of a silicon vertex detector (SVD),
%a 50-layer central drift chamber, an array of
%aerogel threshold Cherenkov counters,
%a barrel-like arrangement of time-of-flight
%scintillation counters,
%an electromagnetic calorimeter (ECL) and an iron flux-return
%instrumented to detect $K_L^0$ mesons and to identify
%muons.
%A 2.0\,cm radius
%beampipe
%and a 3-layer
%SVD
%(SVD1) were used for 
%a 140 fb$^{-1}$ data sample
%containing $152\times 10^6$ $B\bbar$ pairs,
%while a 1.5\,cm radius beampipe, a 4-layer
%silicon detector (SVD2)~\cite{Ushiroda} 
%and a small-cell inner drift chamber were used for 
%an additional 113 fb$^{-1}$
%data sample
%that contains $123\times 10^6$ $B\bbar$
%pairs for a total of $275\times 10^6$ $B\bbar$ pairs.

The Belle detector~\cite{Belle} is a large-solid-angle magnetic
spectrometer. A 2.0\,cm radius beam pipe and a 3-layer silicon vertex
detector (SVD1) were used for a $140\,fb^{-1}$ data sample containing
$152\times 10^6$ $B\bbar$ pairs, while a 1.5\,cm radius beampipe,
a 4-layer silicon detector (SVD2)~\cite{Ushiroda} and a small-cell inner
drift chamber were used for an additional $113\,fb^{-1}$ data sample
that contains $123\times 10^6$ $B\bbar$
pairs for a total of $275\times 10^6$ $B\bbar$ pairs.

%%%%%%%%%%%%%%%%%%%%%%%%%%%%%%%%%%%%%%%%%%%%%%%%%%%%%%%%%%%%%%%%%%%%
%\section{Event Selection, Flavor Tagging and Vertex Reconstruction}
%%%%%%%%%%%%%%%%%%%%%%%%%%%%%%%%%%%%%%%%%%%%%%%%%%%%%%%%%%%%%%%%%%%%
%%%%%%%%%%%%%%%%%%%%%%%%%%%%%%%%%%
%\subsection{Event Selection}
%%%%%%%%%%%%%%%%%%%%%%%%%%%%%%%%%%

For high energy prompt photons,
we select an isolated cluster in the
%ECL
electromagnetic calorimeter (ECL)
that has no corresponding
charged track, and has the largest energy in the c.m.s. 
We require
the shower shape to be consistent with that of a photon.
In order to reduce the background from $\piz$ or $\eta$, 
we exclude photons compatible with $\piz\to\gamma\gamma$ or 
$\eta\to\gamma\gamma$ decays;
we reject photon pairs that satisfy
$\mathcal{L}_{\piz}\ge 0.18$ or $\mathcal{L}_{\eta}\ge 0.18$,
where $\mathcal{L}_{\piz(\eta)}$ is a 
$\piz$ ($\eta$) likelihood described in detail
elsewhere~\cite{Koppenburg:2004fz}.
The polar angle of the
photon direction
in the laboratory frame
is restricted
to the barrel 
region of the ECL ($33^\circ < \theta_\gamma < 128^\circ$),
but is extended to 
the end-cap regions ($17^\circ < \theta_\gamma < 150^\circ$) 
for the second data sample 
due to the reduced material in front of the ECL. 

Neutral kaons ($\ks$) are reconstructed from two oppositely charged pions
that have an invariant mass 
within $\pm 6\MeVcc$ ($2\sigma$) of the $\ks$
nominal mass.
The $\pip\pim$ vertex is required to be displaced from the
interaction point (IP) in the direction of the pion pair
momentum~\cite{Abe:2004xp}. 
Neutral pions ($\piz$) are formed from two photons
with the invariant mass 
within $\pm 16\MeVcc$ ($3\sigma$) of the $\piz$ mass.
The photon momenta are then recalculated with a $\piz$ mass
constraint and we require
the momentum of $\piz$ candidates 
in the laboratory frame to be greater
than $0.3\GeVc$.
The $\ks\piz$ invariant mass, $M_{\ks\piz}$, is required to be
less than 1.8$\GeVcc$. 

$\bz$ mesons are reconstructed by combining $\ks$, $\piz$ and $\gamma$
candidates. We form two kinematic variables:
the energy difference $\dE\equiv E_B^{\rm c.m.s.}-E_{\rm
beam}^{\rm c.m.s.}$ and the beam-energy constrained mass
$\mb\equiv\sqrt{(E_{\rm beam}^{\rm c.m.s.})^2-(p_B^{\rm c.m.s.})^2}$,
where $E_{\rm beam}^{\rm c.m.s.}$ is the beam energy,
$E_B^{\rm c.m.s.}$ and $p_B^{\rm c.m.s.}$
are the energy and the momentum of the candidate in the c.m.s.
Candidates are accepted if they have $\mb > 5.2\GeVcc$ and $|\dE| <
0.5\GeV$.

We reconstruct $\bp\to\ks\pip\gamma$ candidates in a similar
way as the $\bz\to\ks\piz\gamma$ decay
in order to reduce the cross-feed
background from $\bp\to\ks\pip\gamma$ in
$\bz\to\ks\piz\gamma$.
The $\bp\to\ks\pip\gamma$ 
events are also used for various cross-checks.
For a $\pip$ candidate, we require that the track originates from the IP
and that the transverse momentum is greater than $0.1\GeVc$. 
We also require that the $\pip$ candidate cannot be identified as
any other particle species ($K^+, p^+, e^+,$ and $\mu^+$). 

Candidate $\bp\to\ks\pip\gamma$ and
$\bz\to\ks\piz\gamma$ decays are selected simultaneously;
we allow only one candidate for each event.
The best candidate selection is based on the event likelihood ratio
$\rsigbkg$ that is obtained by
combining a Fisher discriminant $\calf$~\cite{Fisher}, which uses
the extended modified Fox-Wolfram moments~\cite{Abe:2003yy} as
discriminating variables, 
and $\cos\theta_H$ defined as the
angle between the $B$ meson momentum and the daughter
$\ks$ momentum in the rest frame of the $\ks\pi$ system.
We select the candidate that has the largest $\rsigbkg$.
For the exclusive $B^0\to\kstarz\gamma$ study, we further require 
$0.8\GeVcc < M_{\ks\piz} < 1.0\GeVcc$ after the best
candidate selection.
The signal region
is defined as 
$-0.2\GeV < \dE < 0.1\GeV$ and $5.27\GeVcc < \mb < 5.29\GeVcc$.

We use events outside the signal region as well as large Monte Carlo
(MC) samples to study the background components.  The dominant
background is from
continuum light quark pair production ($e^+e^-\to q\,\bar{q}$
with 
$q = u,d,s,c$), which we refer to as $\qq$ hereafter.
We require $\rsigbkg > 0.5$ to reduce the $\qq$ background;
after applying all other selection criteria, 
this rejects 79\% of the $\qq$ background
while retaining 88\% of the signal.
Background contributions from $B$ decays,
which are considerably smaller than $\qq$,
are dominated by cross-feed from 
other radiative $B$ decays including
$\bp\to\ks\pip\gamma$.

%%%%%%%%%%%%%%%%%%%%%%%%%%
%\subsection{Flavor Tagging}
%\label{sec:flavor tagging}
%%%%%%%%%%%%%%%%%%%%%%%%%%%

The $b$-flavor of the accompanying $B$ meson is identified
from inclusive properties of particles
that are not associated with the reconstructed signal
decay.
The algorithm for flavor tagging is described in detail
elsewhere~\cite{bib:fbtg_nim}.
We use two parameters, $\fq$ defined in Eq.~(\ref{eq:psig}) and $r$, to
represent the tagging information. 
The parameter $r$ is an event-by-event
flavor-tagging
dilution factor that ranges from 0 to 1; $r=0$
when there is no flavor 
discrimination and $r=1$
implies unambiguous flavor assignment. 
It is
determined by using MC data and is only used 
to sort data into six $r$ intervals.
The wrong tag fraction $w$
and the difference $\Delta w$ in $w$
for the $\bz$ and $\bzb$ decays
are determined for each of the six $r$ intervals
from data~\cite{Abe:2004xp}.

%%%%%%%%%%%%%%%%%%%%%%%%%%%%%%%%
%\subsection{Vertex Reconstruction}
%%%%%%%%%%%%%%%%%%%%%%%%%%%%%%%%

The vertex position of the signal-side decay is
reconstructed from the $\ks$ trajectory with a constraint on the IP;
the IP profile
($\sigma_x\simeq 100\rm\,\mu m$,
$\sigma_y\simeq 5\rm\,\mu m$,
$\sigma_z\simeq 3\rm\,mm$)
is convolved with
the
finite $B$ flight length in the plane
perpendicular to the $z$ axis.
Both pions from the $\ks$ decay are required to
have enough hits
in the SVD in order to reconstruct the $\ks$
trajectory with high resolution.
The reconstruction efficiency depends not only on
the $\ks$ momentum but also on the SVD geometry.
The efficiency with SVD2 (55\%) is
higher than with SVD1 (41\%)
because of the larger
detection volume.
The other (tag-side) $B$ vertex determination is the same as that for
the $\bz\to\phi\ks$ analysis~\cite{Abe:2004xp}.

%%%%%%%%%%%%%%%%%%%%%%%%%%%%%%%%%%%%%%%%%%%%%%%%%%%%%%%%%%%%%%%%
% signal yield extraction
%%%%%%%%%%%%%%%%%%%%%%%%%%%%%%%%%%%%%%%%%%%%%%%%%%%%%%%%%%%%%%%%

\begin{figure}
\resizebox{0.46\columnwidth}{!}{\includegraphics{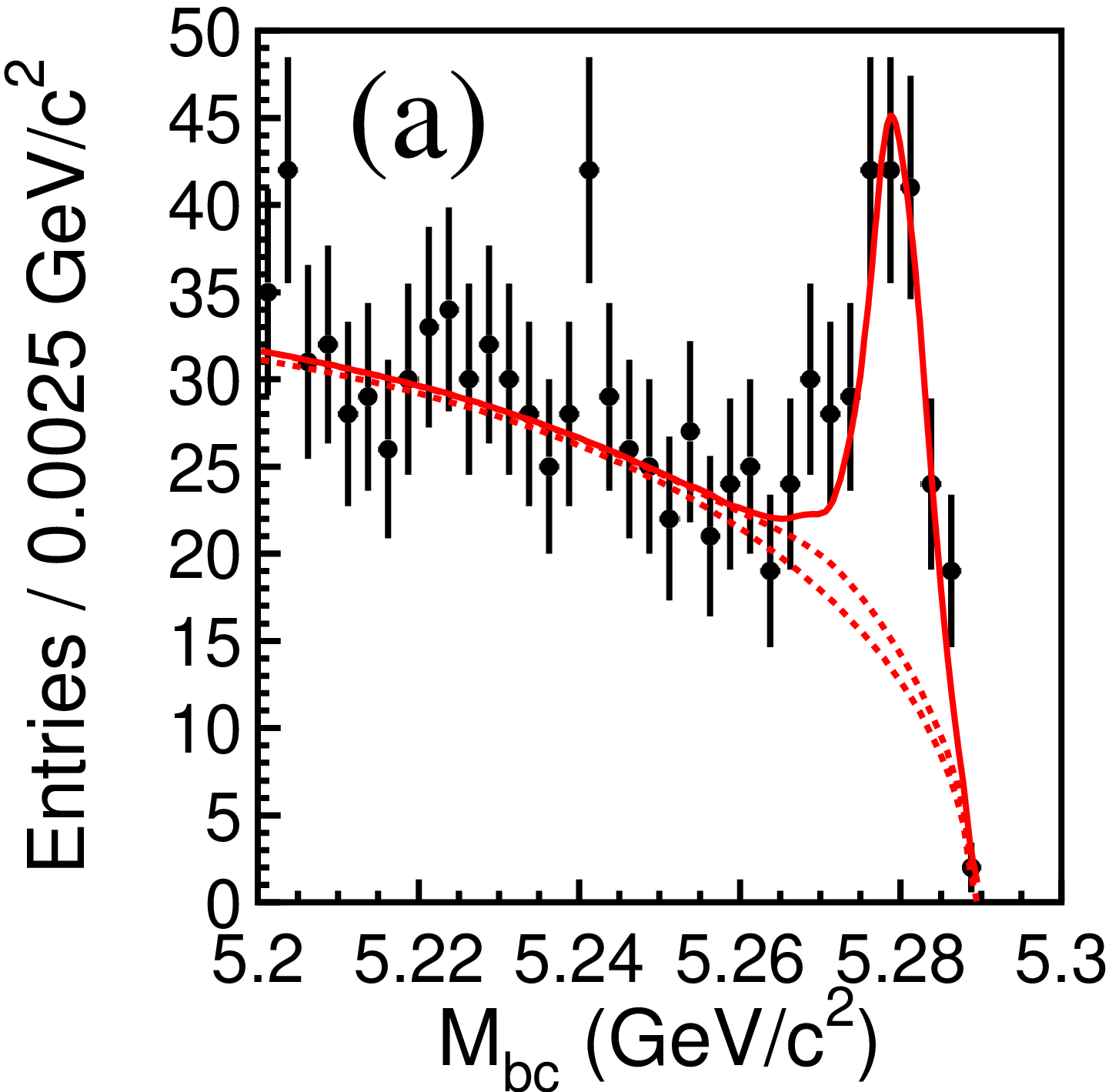}}
\resizebox{0.46\columnwidth}{!}{\includegraphics{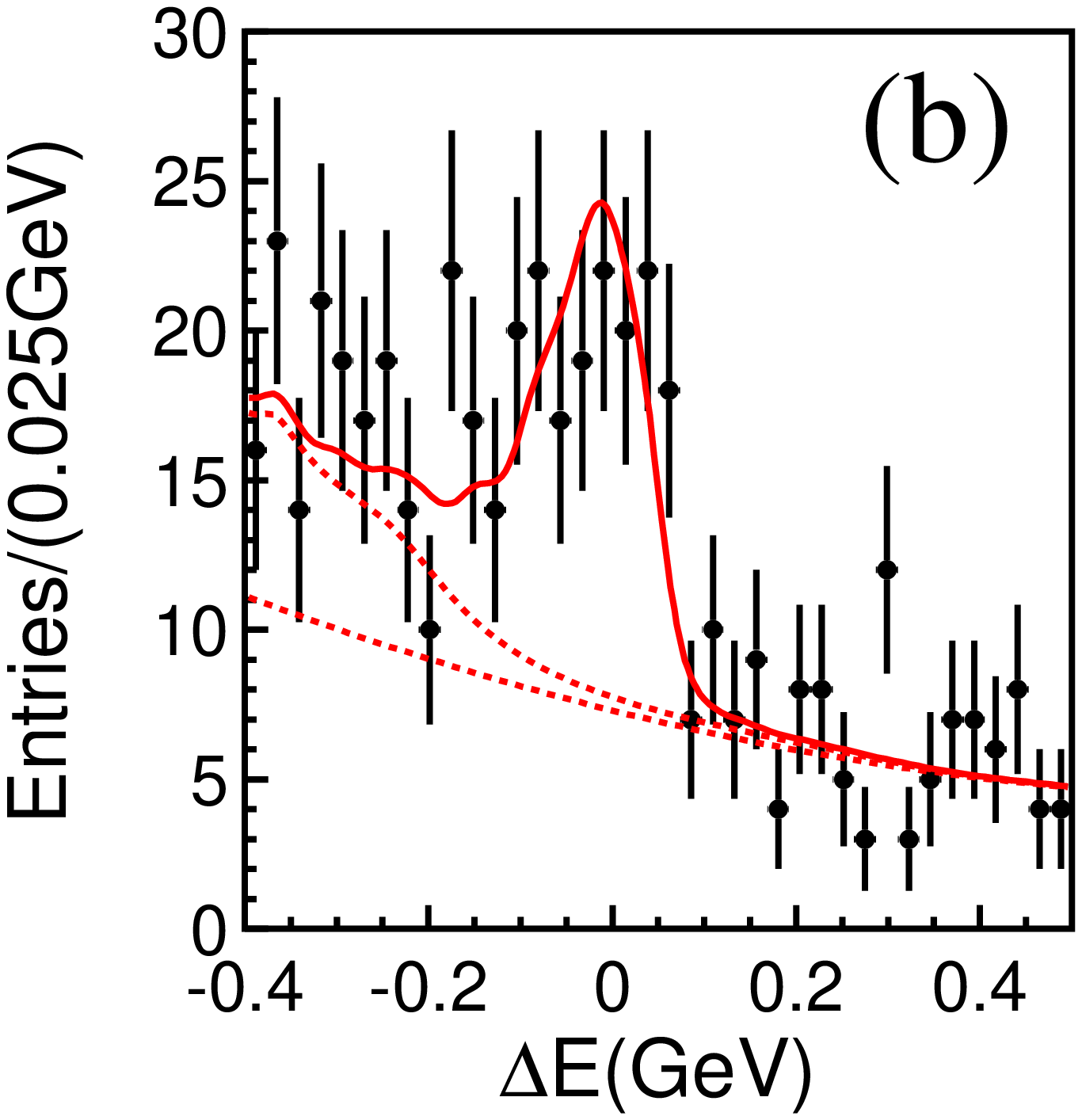}}
\caption{(a) $\mb$ distributions within the $\dE$ signal region
and (b) $\dE$ distributions within the $\mb$ signal region for
$\bz\to\ks\piz\gamma$.
Solid curves show the fit to signal plus background distributions.
Lower (upper) dashed curves show the background contributions from $\qq$
 ($\qq$ and $B$ decays).
 }
\label{fig:mb}
\end{figure}

\begin{figure}
\resizebox{0.7\columnwidth}{!}{\includegraphics{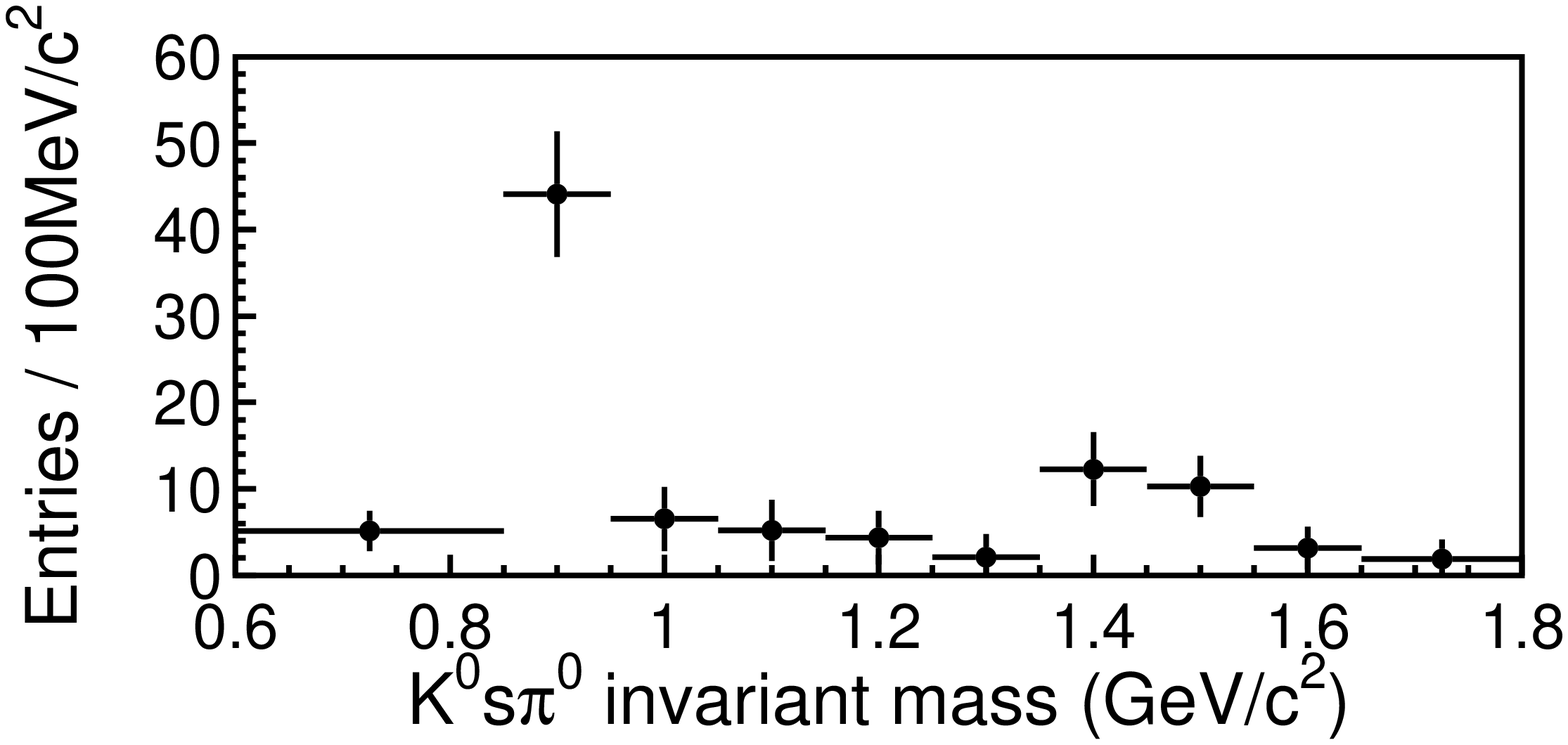}}
\caption{$M_{\ks\piz}$ distribution of the
 $\bz\to\ks\piz\gamma$ signal yield obtained from the fit to the
 $\dE$-$\mb$ distribution.
 Background shape parameters are fixed at the values obtained with
 the whole $M_{\ks\piz}$ region.
 }
\label{fig:mkspiz}
\end{figure}

Figure~\ref{fig:mb} shows the $\mb$ ($\dE$) distribution for the
reconstructed $\ks\piz\gamma$ candidates within the $\dE$ ($\mb$)
signal region after flavor tagging and vertex reconstruction.  The
signal yield is determined from an unbinned two-dimensional
maximum-likelihood fit to the $\dE$-$\mb$ distribution. The fit region
for the $\bz\to\kstarz\gamma$ analysis is $ 5.20\GeVcc < \mb <
5.29\GeVcc$ and
$-0.5\GeV < \dE < 0.5\GeV$. For the $\bz\to\ks\piz\gamma$ analysis, the
$\dE$ fit region is narrowed to $-0.4\GeV < \dE < 0.5\GeV$ in order to
avoid other $B$ background events that populate
the low $\dE$ region. 
The signal distribution is represented by a smoothed histogram obtained
from MC simulation that accounts for a
small correlation
between $\mb$ and $\dE$.  The background from $B$ decays is also modeled
with a smoothed histogram obtained from MC events.  For the $\qq$
background, we use the ARGUS parameterization~\cite{bib:ARGUS} for $\mb$
and a second-order
polynomial 
for $\dE$.  Normalizations of the
signal and background distributions and the $\qq$ background shape are
the five free parameters in the fit.
We observe 501 (215) $\ks\piz\gamma$ ($\kstarz\gamma$) candidates in the
signal box, which decrease to 227 (92) after the flavor tagging and
the $B$ vertex reconstruction, and obtain $\Nsigkspizgm$ ($\Nsigkstarzgm$) 
signal events by the fit.
Figure~\ref{fig:mkspiz} shows
the signal yields in different 
$M_{\ks\piz}$ regions. For each bin, a fit to the $\dE$-$\mb$ distribution
is performed, and the obtained signal yield is
plotted. 

%%%%%%%%%%%%%%%%%%%%%%%%%%%%%%%%%%%%%%%%%%%%%%%%%
%\section{Results of {\boldmath $CP$} Asymmetry Measurements}
%%%%%%%%%%%%%%%%%%%%%%%%%%%%%%%%%%%%%%%%%%%%%%%%%

We determine $\cals$ and $\cala$
from an unbinned
maximum-likelihood fit to the observed $\Dt$ distribution.
The probability density function (PDF) expected for the signal
distribution,
${\cal P}_{\rm sig}(\Dt;\cals,\cala,\fq,w,\Delta w)$, 
is given by the time dependent decay rate [Eq.~(\ref{eq:psig})]
modified to incorporate 
the effect of incorrect flavor assignment. The distribution is
convolved with the
proper-time interval resolution function 
$\Rsig$, 
which takes into account the finite vertex resolution. 
The parametrization of $\Rsig$ is the same as that used for the
$\bz\to\ks\piz$ decay~\cite{Abe:2004xp}. $\Rsig$ is first derived from
flavor-specific $B$ decays~\cite{bib:BELLE-CONF-0436} and
modified by additional parameters which rescale vertex errors to
account for the fact that there is no track directly originating from
the IP. 

For each event, the following likelihood function is
evaluated:
\begin{eqnarray}
P_i
&=& (1-\fol)\int_{-\infty}^{+\infty} \biggl[
\fsig{\cal P}_{\rm sig}(\Dt')\Rsig (\Dt_i-\Dt') \nonumber \\
&+&(1-\fsig){\cal P}_{\rm bkg}(\Dt')\Rbkg (\Dt_i-\Dt')\biggr]
d(\Dt')  \nonumber \\
&+&\fol P_{\rm ol}(\Dt_i),
\label{eq:likelihood}
\end{eqnarray}
where $P_{\rm ol}$ 
is a Gaussian function that represents a 
small outlier component with fraction $\fol$~\cite{bib:resol}. 
The signal probability $\fsig$ is calculated on an event-by-event basis
from the function which we obtained as the result of the
two-dimensional $\dE$-$\mb$ fit for the signal yield
extraction. 
A PDF for background events,
${\cal P}_{\rm bkg}$, 
is modeled as a sum of exponential and prompt components, and
is convolved with a sum of two Gaussians
which represent the resolution function $\Rbkg$ for the
background. 
All parameters in
${\cal P}_{\rm bkg}$ 
and $\Rbkg$ are determined by
a 
fit to the $\Dt$ distribution of a 
background-enhanced control sample, i.e. events 
outside of the
$\dE$-$\mb$ signal region.
We fix $\tau_\bz$ and $\dmd$ at
their world-average values~\cite{bib:PDG2004}.

The only free parameters in the final fit for the
$\bz\to\ks\piz\gamma$ decay
are $\cals_{\ks\piz\gamma}$ and $\cala_{\ks\piz\gamma}$, which are
determined by maximizing the likelihood function
$L = \prod_iP_i(\Dt_i;\cals,\cala)$
where the product is over all events.
For the $\bz\to\kstarz\gamma$ decay, we fix $\cala_{\kstarz\gamma}$ at
zero and perform a single parameter fit for $\cals_{\kstarz\gamma}$.
This is based on the previous measurement~\cite{bib:NakaoAcp} of the
partial rate asymmetry in $B\to\kstar\gamma$ decays,
which yielded $\cala_{\kstarz\gamma}$ consistent with zero. 
The error of the previous measurement
of $\cala_{\kstarz\gamma}$
is included in 
the systematic error of $\cals_{\kstarz\gamma}$.
We obtain
\begin{eqnarray}
 \cals_{\kstarz\gamma} &=& \SkstarzgmResultSS,\nonumber\\
 \cals_{\ks\piz\gamma} &=& \SkspizgmResultSS,\nonumber \\
 \cala_{\ks\piz\gamma} &=& \AkspizgmResultSS.\nonumber
\end{eqnarray}

\begin{figure}
\resizebox{!}{0.46\columnwidth}{\includegraphics{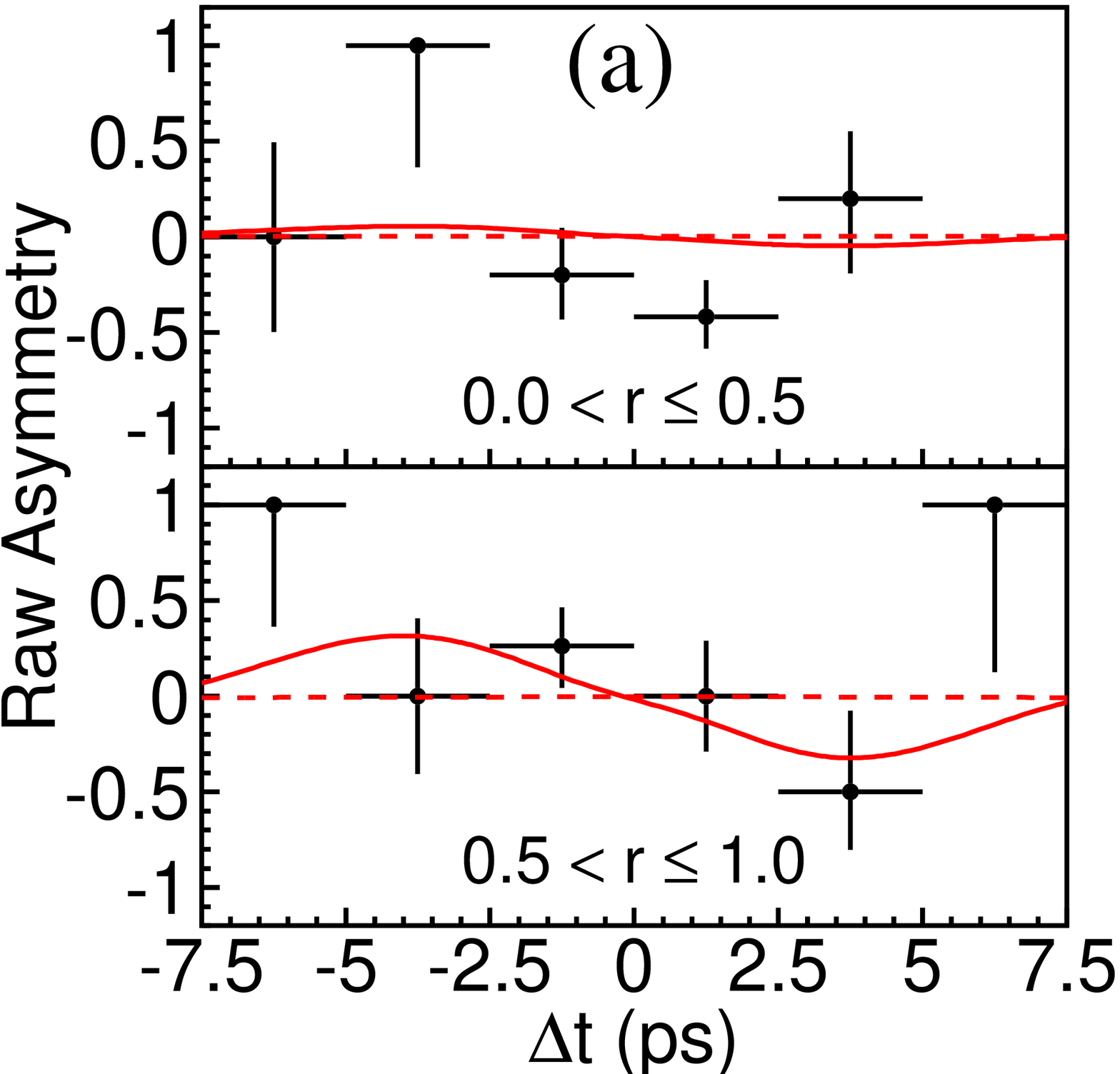}} 
\resizebox{!}{0.46\columnwidth}{\includegraphics{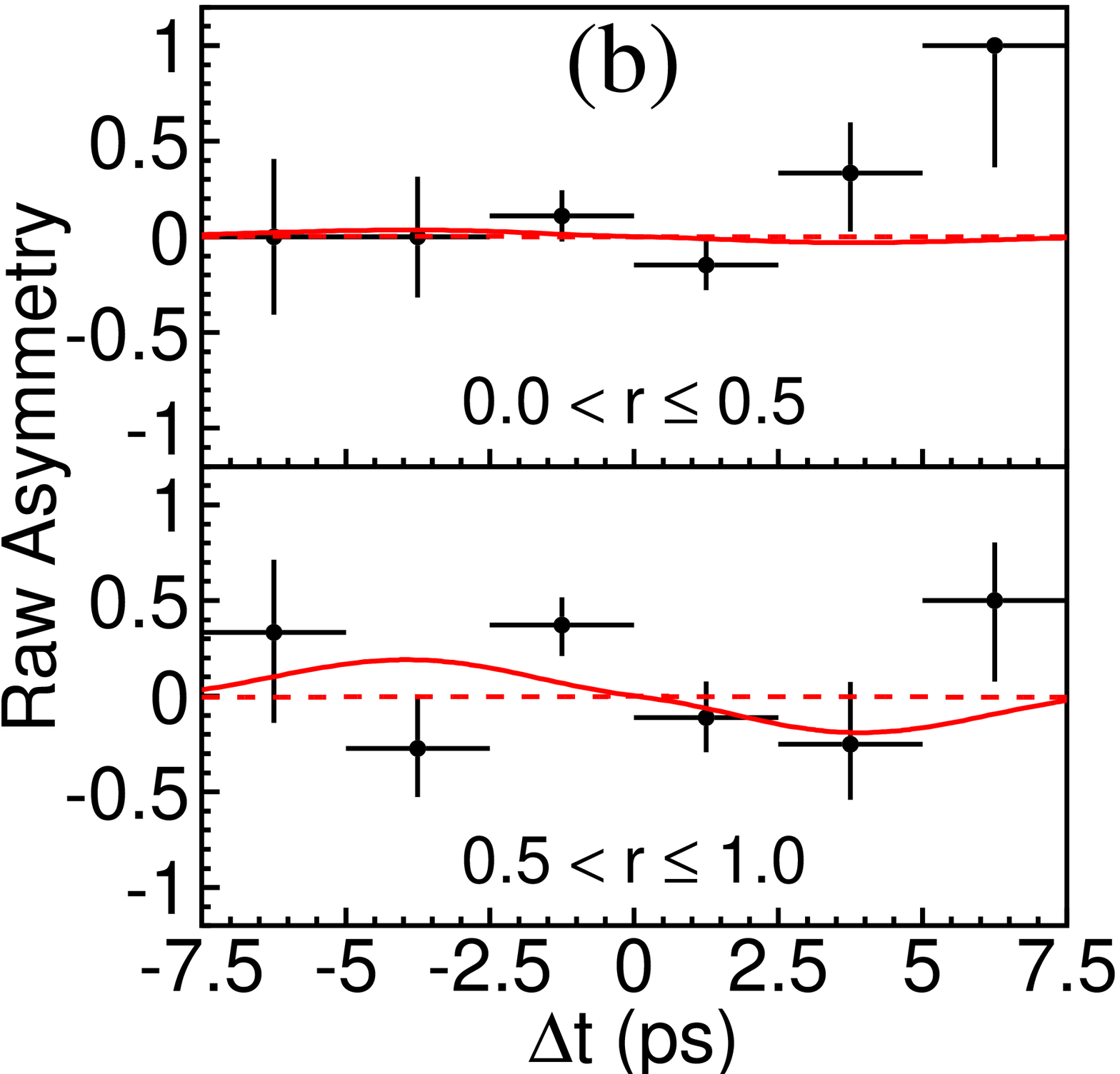}} 
\caption{
 Asymmetry in each $\Dt$ bin with $0 < r \le 0.5$ (top)
 and with $0.5 < r \le 1.0$ (bottom) for (a) $\bz\to\kstarz\gamma$ and
 for (b) $\bz\to\ks\piz\gamma$.
 Solid curves show the results of the unbinned maximum-likelihood
 fits.
 The dashed curves show the null asymmetry case ($\cals=\cala=0$).
 The dilution of a possible asymmetry is smaller for the $0.5 < r \le
 1.0$ subsample, because of the lower probability of incorrect flavor
 assignment.
 }
\label{fig:asym}
\end{figure}

We define the raw asymmetry in each $\Dt$ bin by
$(N_{q=+1}-N_{q=-1})/(N_{q=+1}+N_{q=-1})$,
where $N_{q=+1(-1)}$ is the number of 
observed candidates with $q=+1(-1)$.
Figure~\ref{fig:asym}
shows the raw asymmetries for the
$\ks\piz\gamma$ and $\kstarz\gamma$ decays
in two regions of the flavor-tagging
parameter $r$.
Note that these are simple projections onto the $\Delta t$ axis, and do
not reflect other event-by-event information (such as the signal
fraction, the wrong tag fraction and the vertex resolution), which is in
fact used in the unbinned maximum-likelihood fit for $\cals$ and
$\cala$.

%%%%%%%%%%%%%%%%%%%%%%%%%%%
%\subsection{Systematic Error}
%%%%%%%%%%%%%%%%%%%%%%%%%%%

Primary sources of the systematic error are
(1) uncertainties in the resolution function
($\pm 0.05$ for $\cals_{\ks\piz\gamma}$,
  $\pm 0.04$ for $\cala_{\ks\piz\gamma}$ and
  $\pm 0.05$ for $\cals_{\kstarz\gamma}$),
(2) uncertainties in the vertex reconstruction 
($\pm 0.05$ for $\cals_{\ks\piz\gamma}$,
  $\pm 0.06$ for $\cala_{\ks\piz\gamma}$ and
  $\pm 0.06$ for $\cals_{\kstarz\gamma}$),
(3) uncertainties in the background fraction
($\pm 0.06$ for $\cals_{\ks\piz\gamma}$,
  $\pm 0.04$ for $\cala_{\ks\piz\gamma}$ and
  $\pm 0.02$ for $\cals_{\kstarz\gamma}$),
and (4) uncertainties in the background $\Dt$ distribution
($\pm 0.05$ for $\cals_{\ks\piz\gamma}$,
  $\pm 0.04$ for $\cala_{\ks\piz\gamma}$ and
  $\pm 0.02$ for $\cals_{\kstarz\gamma}$).
Effects of tag-side interference~\cite{Long:2003wq}
contribute
$\pm 0.01$ for $\cals_{\ks\piz\gamma}$,
$\pm 0.06$ for $\cala_{\ks\piz\gamma}$ and
$\pm 0.01$ for $\cals_{\kstarz\gamma}$.
Also included are effects 
from uncertainties in the wrong tag fraction and
physics parameters ($\dmd$, $\taubz$ and $\cala_{\kstarz\gamma}$). 
A possible bias in the fit is estimated by using MC data. 
The total systematic error is obtained by adding
these contributions in quadrature.

%%%%%%%%%%%%%%%%%%%%%%
%\section{Cross Check}
%%%%%%%%%%%%%%%%%%%%%%

Various cross-checks of the measurement are performed.
We apply the same fit procedure to the $\bz\to\jpsi\ks$ sample
without using $\jpsi$ daughter tracks for the vertex
reconstruction. We obtain
$\cals_{\jpsi\ks} = +0.68\pm 0.10$(stat) and
$\cala_{\jpsi\ks} = +0.02\pm 0.04$(stat), which
are in good agreement with the world average
values~\cite{bib:PDG2004}.
We perform a fit
to $\bp\to\ks\pip\gamma$
($\bp\to\kstarp\gamma\to\ks\pip\gamma$), 
which is a good control sample of the $\bz\to\ks\piz\gamma$
($\bz\to\kstarz\gamma\to\ks\piz\gamma$) 
decay, without using the primary $\pip$ for the vertex reconstruction. 
The result is consistent with no $CP$ asymmetry, as expected. 
Lifetime measurements are also performed for these modes, and 
values consistent with the world-average are obtained. 
Ensemble tests are
carried out 
with MC pseudo-experiments;
we find that the statistical errors obtained
in our measurements are all within the expectations from the ensemble
tests.
A fit to the $\bz\to\ks\piz\gamma$ subsample 
that 
excludes the $\kstarz$ mass region
yields
$\cals = -0.39^{+0.63}_{-0.52}$(stat) and
$\cala = +0.10\pm 0.51$(stat).
The result is consistent with the results for $\bz\to\kstarz\gamma$
and for the full $\bz\to\ks\piz\gamma$ sample.

%%%%%%%%%%%%%%%%%%
%\section{Summary}
%%%%%%%%%%%%%%%%%%

In summary, 
we have performed a new measurement of the time-dependent $CP$ asymmetry
in the decay $\bz\to\ks\piz\gamma$ based on a sample of $275\times
10^6$ $B\bbar$ pairs.
Two regions of the $\ks\piz$ invariant mass are used:
between $0.8\GeVcc$ and $1.0\GeVcc$ for the $\kstarz$ resonance,
and the full region up to $1.8\GeVcc$ 
including other resonances and non-resonant contributions.
We obtain $CP$-violation parameters in the $\bz\to\ks\piz\gamma$ decay
$\cals_{\ks\piz\gamma}=\SkspizgmResultSS$ and
$\cala_{\ks\piz\gamma}=\AkspizgmResultSS$,
and in
$\bz\to\kstarz(\to\ks\piz)\gamma$ decay
$\cals_{\kstarz\gamma}=\SkstarzgmResultSS$ with
$\cala_{\kstarz\gamma}$ fixed at zero.
The value of $\cals_{\kstarz\gamma}$ is consistent with the measurement by
\BaBar~\cite{Aubert:2004pe}.
$CP$-violation parameters in the three-body $\bz\to\ks\piz\gamma$
decay are measured for the first time.
The two results 
are consistent with each other and with no $CP$ asymmetry,
while the statistical error is
considerably
reduced by the inclusion of these additional events in the
$\bz\to\ks\piz\gamma$ analysis. 

%%%%%%%%%%%%%%%%%%%%%%%%%%%
%\section*{Acknowledgments}
%%%%%%%%%%%%%%%%%%%%%%%%%%%

%ack.tex
%-------- Short version, if necessary, for PRL -----------
We thank the KEKB group for the excellent operation of the
accelerator, the KEK cryogenics group for the efficient
operation of the solenoid, and the KEK computer group and
the NII for valuable computing and Super-SINET network
support.  We acknowledge support from MEXT and JSPS (Japan);
ARC and DEST (Australia); NSFC (contract No.~10175071,
China); DST (India); the BK21 program of MOEHRD and the CHEP
SRC program of KOSEF (Korea); KBN (contract No.~2P03B 01324,
Poland); MIST (Russia); MHEST (Slovenia);  SNSF (Switzerland); NSC and MOE
(Taiwan); and DOE (USA).

%\input{ack.tex}

%%%%%%%%%%%%%%%%%%%%%%%%%%
% Bibliography
%%%%%%%%%%%%%%%%%%%%%%%%%%


\begin{thebibliography}{999}

\bibitem{Kobayashi:1973fv}
M.~Kobayashi and T.~Maskawa, Prog.\ Theor.\ Phys. {\bf 49}, 652 (1973).

%\bibitem{bib:sanda}
%A.~B.~Carter and A.~I.~Sanda, Phys.\ Rev.\ D \textbf{23}, 1567 (1981);
%I.~I.~Bigi and A.~I.~Sanda, Nucl.\ Phys. \textbf{B193}, 85 (1981).

\bibitem{Atwood:1997zr}
%$\cals$ is suppressed by a factor of $-2(m_s/m_b)$, according to
D.~Atwood, M.~Gronau and A.~Soni,
%``Mixing-induced CP asymmetries in radiative B decays in and beyond the
%standard model,''
Phys.\ Rev.\ Lett.\  {\bf 79}, 185 (1997).
%[arXiv:hep-ph/9704272].
%%CITATION = HEP-PH 9704272;%%

\bibitem{Grinstein:2004uu}
%$\cals \sim 0.1$ is allowed within the SM according to
B.~Grinstein, Y.~Grossman, Z.~Ligeti and D.~Pirjol,
%``The photon polarization in B $\to$ X gamma in the standard model,''
Phys.\ Rev.\ D {\bf 71}, 011504 (2005).
%[arXiv:hep-ph/0412019].
%%CITATION = HEP-PH 0412019;%%

\bibitem{Atwood:2004jj}
D.~Atwood, T.~Gershon, M.~Hazumi and A.~Soni,
%``Mixing-induced CP violation in B $\to$ P(1) P(2) gamma in search of clean
%new physics signals,''
%hep-ph/0410036.
%%CITATION = HEP-PH 0410036;%%
Phys.\ Rev.\ D {\bf 71}, 076003 (2005).

\bibitem{bib:Kstar}
	Throughout this paper,
	$\kstar$ denotes $\kstar(892)$.
	The inclusion of the charge conjugate decay mode is implied.

\bibitem{note:Grinstein}
Possible strong phase dependence within the SM was pointed out by
	Grinstein {\it et al.}~\cite{Grinstein:2004uu}.
In the case $CP$ violation depends on the $\ks\piz$ mass, our measurement
for $\ks\piz\gamma$ gives an efficiency-weighted average of $CP$
asymmetries over the extended mass range.

\bibitem{bib:KEKB}
S.~Kurokawa and E.~Kikutani, %{\it et al.}, 
Nucl. Instr. and Meth. A {\bf 499}, 1 (2003).

\bibitem{Belle}
Belle Collaboration, A.~Abashian {\it et al.},
Nucl. Instr. and Meth. A {\bf 479}, 117 (2002).

\bibitem{Ushiroda} Y.~Ushiroda (Belle SVD2 Group),
Nucl. Instr. and Meth. A {\bf 511}, 6 (2003).

\bibitem{Koppenburg:2004fz}
Belle Collaboration, P.~Koppenburg {\it et al.},
%``An inclusive measurement of the photon energy spectrum in b $\to$ s gamma
%decays,''
Phys.\ Rev.\ Lett.\  {\bf 93}, 061803 (2004).
%[arXiv:hep-ex/0403004].
%%CITATION = HEP-EX 0403004;%%

\bibitem{Abe:2004xp}
Belle Collaboration, K.~Abe {\it et al.},
%``New measurements of time-dependent CP-violating asymmetries in b $\to$ s
%transitions at Belle,''
hep-ex/0409049.
%%CITATION = HEP-EX 0409049;%%

\bibitem{Fisher}
R.~A.~Fisher, Annals Eugen. {\bf 7}, 179 (1936).

%\cite{Abe:2003yy}
\bibitem{Abe:2003yy}
Belle Collaboration, K.~Abe {\it et al.},
%``Evidence for B0 $\to$ pi0 pi0,''
Phys.\ Rev.\ Lett.\  {\bf 91}, 261801 (2003).
%[arXiv:hep-ex/0308040].
%%CITATION = HEP-EX 0308040;%%

\bibitem{bib:fbtg_nim}
H.~Kakuno, K.~Hara {\it et al.},
Nucl. Instr. and Meth. A {\bf 533}, 516 (2004).
%hep-ex/0403022, to appear in Nucl. Instr. and Meth. A.

\bibitem{bib:ARGUS}
ARGUS Collaboration, H.~Albrecht \textit{et al.}, 
Phys. Lett. B {\bf 241}, 278 (1990).

\bibitem{bib:BELLE-CONF-0436}
Belle Collaboration, K.~Abe {\it et al.}, 
%BELLE-CONF-0436, U0202
hep-ex/0408111.

\bibitem{bib:resol}
Belle Collaboration, K.~Abe {\it et al.},
% ``Improved measurement of CP-violation parameters sin(2phi(1)) and
%$|$lambda$|$, B meson lifetimes, and B0 anti-B0 mixing parameter
%Delta(m(d)),''
Phys.\ Rev.\ D {\bf 71}, 072003 (2005)
%[Erratum-ibid.\ D {\bf 71}, 079903 (2005)]
%[arXiv:hep-ex/0408111].
%%CITATION = HEP-EX 0408111;%%
;
H.~Tajima {\it et al.},
Nucl.\ Instrum.\ Methods Phys. Res., Sect. A {\bf 533}, 370 (2004).

\bibitem{bib:PDG2004}
S.~Eidelman {\it et al.}, Phys. Lett. B {\bf 592}, 1 (2004).

%%\cite{Nakao:2004th}
\bibitem{bib:NakaoAcp}
Belle Collaboration, M.~Nakao {\it et al.},  
%%``Measurement of the B $\to$ K* gamma branching fractions and asymmetries,''
Phys.\ Rev.\ D {\bf 69}, 112001 (2004).
%%[arXiv:hep-ex/0402042].
%%CITATION = HEP-EX 0402042;%%

\bibitem{Long:2003wq}
O.~Long, M.~Baak, R.~N.~Cahn and D.~Kirkby,
%``Impact of tag-side interference on time dependent CP asymmetry  measurements
%using coherent B0 anti-B0 pairs,''
Phys.\ Rev.\ D {\bf 68}, 034010 (2003).
%[arXiv:hep-ex/0303030].
%%CITATION = HEP-EX 0303030;%%

\bibitem{Aubert:2004pe}
%B.~Aubert {\it et al.} [\BaBar\ Collaboration],
\BaBar\ Collaboration, B.~Aubert {\it et al.},
% ``Measurement of time-dependent CP-violating asymmetries in B0 $\to$ K*0 gamma ``
%(K*0 $\to$ K0(S) pi0) decays,''
Phys.\ Rev.\ Lett.\ {\bf 93}, 201801 (2004).
%[arXiv:hep-ex/0405082].
%%CITATION = HEP-EX 0405082; %%	

\end{thebibliography}
\end{document}